\documentclass{elsart}
\usepackage{natbib}
\usepackage{color,graphicx}
\usepackage{amsmath,amssymb,amsfonts}

\newcommand{\grad}{\boldsymbol{\nabla}} 
\newcommand{\di}{\partial} 
\newcommand{\Fvec}{{\bf F}} 

\begin{document}

\begin{frontmatter}

\title{A fully 3-dimensional thermal model of a comet nucleus}

\author{Eric D Rosenberg}
\ead{erosenbe@post.tau.ac.il}
\ead[url]{http://geophysics.tau.ac.il/personal/erosenbe}
and
\author{Dina Prialnik}
\ead{dina@planet.tau.ac.il}
\address{Department of Geophysics and Planetary Sciences,\\ Raymond and Beverly Sackler Faculty of Exact Sciences,\\ Tel-Aviv University,\\ Tel-Aviv 69978, Israel.}

\address{}

\begin{abstract}
A 3-D numerical model of comet nuclei is presented. An implicit numerical scheme was developed for the thermal evolution 
of a spherical nucleus composed of a mixture of ice and dust. The model was tested against analytical solutions, simplified numerical
solutions, and 1-D thermal evolution codes. The 3-D code was applied to comet 67P/Churyumov-Gerasimenko; surface temperature maps and
the internal thermal structure was obtained as function of depth, longitude and hour angle. The effect of the spin axis tilt on the
surface temperature distribution was studied in detail. It was found that for small tilt angles, relatively low temperatures 
may prevail on near-pole areas, despite lateral heat conduction. A high-resolution run for a comet model of 67P/Churyumov-Gerasimenko
with low tilt angle, allowing for crystallization of amorphous ice, showed that the amorphous/crystalline ice boundary 
varies significantly with depth as a function of cometary latitude.
\end{abstract}


\end{frontmatter}

\section{Introduction}\label{S:SciBack}

Our knowledge of the nature of comets, their structure, composition, dynamical 
history, and modes of activity has greatly improved over the last twenty years. 
The one event that constituted a mile-stone in the study of comets was the space mission {\it Giotto},
which provided the first close-up picture of a comet's heart - its nucleus. Several other
space missions to comets followed, crowned by the very recent {\it Deep Impact} mission \citep{AHearn2005}, aimed at
revealing properties of the nucleus interior. As observational data on comets has been accumulating, 
theoretical studies and modeling have started to develop, in order to interpret, predict and
explain new findings and discoveries.

Comets consist of frozen gases and dust. The dust consists of silicates and organic materials, 
while the frozen gases are mostly water and a few constituents more volatile than water. Solar heat, 
not reflected or radiated from the surface of the nucleus is primarily used to evaporate ices from exposed 
areas. The remainder of the heat penetrates into the nucleus, where it can cause numerous physico-chemical 
reaction processes, which eventually affect measurable properties at the nucleus surface. Over the past 
two decades, numerical studies have attempted to integrate some of the more influential processes with a 
numerical representation of a comet nucleus, in order to better understand the rather unpredictable behavior 
of comets \citep{Prialnik2004}.

The most obvious property of comet nuclei is the {\it lack of} spherical symmetry. 
Besides the fact that self-gravity, which dictates the spherical shape of larger celestial bodies, 
is negligible in comets, the main energy source --- solar radiation --- is unevenly distributed over 
the nucleus surface, not only because of daily variation and spin axis inclination, but also 
because of the large orbital variations,
characteristic to comets, but not to other bodies of the solar system. Nevertheless, theoretical models
of nuclei have so far assumed spherical symmetry, or have 
circumvented this assumption by crude approximations.

Modeling began with very simple pictures of the nucleus. A few physical processes with simplifying 
assumptions provided crude but insightful explanations for the comet nucleus behavior \citep{Cowan1979,Fanale1984,Herman1985,Prialnik1987}. However, these 
models were able to give only average results and explain only the general behavior of comets. The models 
were limited to a single dimension, mostly based on spherical symmetry (the {\it fast rotator} approximation). 
They simulated heat flow in a mixture of ice and rock (dust) with uniform boundary conditions for the 
solar flux, radiative emission and substance sublimation. Additional physico-chemical processes,
e.g., amorphous ice crystallization, surface erosion etc., were added later on to produce a more 
elaborate physical picture.
As these models did not take into consideration spatial variations in surface temperature, 
they provided a fair approximation only for the internal part of the nucleus, where these variations fade. 

For more realistic surface results, 
{\it slow rotator} approaches 
were adopted. 
A "1.5-D" model \citep{Benkhoff1996,Benkhoff1999} was suggested, in which, a 1-D model 
was applied with boundary conditions of an equatorial point on a rotating nucleus. This way, diurnal boundary 
conditions differences were taken into considerations, and an upper limit for production rate was achieved when 
the sub-solar (high noon) flux was adopted for the entire surface of the sunlit hemisphere. This model, as its 
predecessor, does not calculate lateral flow. A semi-lateral approach was taken for the 2.5-D model 
\citep{Enzian1997,Enzian1999} where a meridian flow was calculated, and boundary conditions were taken as in the 
1.5-D model. 

A different approach is the quasi-3-D procedure \citep{Gutierrez2000,Julian2000,Cohen2003}. 
In this approach, every point on the nucleus surface is calculated (with the appropriate local boundary condition) 
with radial flow only, so calculations are similar to the 1.5-D method, but boundary conditions are of the 
total sphere, just as with a 3-D approach. Still, each point (or element) of the surface evolves independently
of the others and only radial conduction is considered.

Now that modeling has made significant progress towards understanding the general structure and behavior 
of comets, more sophisticated and realistic pictures of the nucleus are required.

The purpose of the present research project is to develop a {\it fully} 3-D model of a comet nucleus. 
As in the quasi-3-D, boundary conditions will be taken for the entire sphere, but now, meridional and 
azimuthal heat fluxes are calculated as well. Gas production and flow in the interior of the
comet are not included in the present model. In order to compare production rates of volatiles with the
observed ones at large distances from the nucleus, when the surface is not resolved 
(in fact, not even seen) \citep{Biver1999}, a sum of all local production 
rates can be calculated. However, the model will provide a detailed description of the surface activity as well.

In the 3-D model, in contrast to the older models, all the grid elements need to be solved simultaneously, 
so the computational load increases dramatically. This may limit the spatial resolution of the comet if we 
require a solution in a reasonable time. However, as modern computers advance rapidly, the grid density may increase 
for the same calculation time. This model may be scaled for super-computers to allow even denser grid resolution.

\section{Model description}\label{S:methods}

Our purpose is to calculate numerically a variety of physical processes that are believed to take place 
inside a comet nucleus.
In order to do so, physical equations and models should be adjusted for discrete calculations. The basic adjustment is dividing the nucleus 
into elements via a grid, and simplifying the problem by assuming a ``homogeneous lumped system'' approximation for each element.   

\subsection{3-D heat conduction equation}\label{SS:conduction}

Heat conduction is modeled by Fourier's law
\begin{equation}\label{eq:simplefourier}
\Fvec=-K\grad{T},
\end{equation}
where $\Fvec$\, is the heat flux, $K$ is the conduction coefficient and $T$ is the temperature.\\In 3-D spherical coordinates, equation (\ref{eq:simplefourier}) will take the form
\begin{equation}\label{eq:sphericalfourier}
\Fvec=-H_zK\left( \frac{\di T}{\di r} \,\boldsymbol{\hat{r}} + 
                 \frac{1}{r}\frac{\di T}{\di \theta} \,\boldsymbol{\hat{\theta}} +
		 \frac{1}{r\sin{\theta}}\frac{\di T}{\di \phi} \,\boldsymbol{\hat{\phi}}\right)
\end{equation}
where $H_z$ is the Hertz factor, a correction factor ($<1$) for the heat conductivity of porous substances that accounts for
the reduced contact area between solids as compared to the cross-sectional area.
In order to calculate the effective conduction coefficient $K$ for a ``homogeneous lumped system'' we considered all substances 
distributed evenly through the system, and the contribution of each substance to the final value is taken to be
proportional to its mass fraction $X_\alpha$. Thus
\begin{equation}
K=\sum_{\alpha}{X_\alpha K_\alpha},
\end{equation}
where $K_\alpha$ is the conduction coefficient of substance $\alpha$.

When all fluxes through a ``lumped system'' are combined we get the heat equation
\begin{equation}\label{eq:heat}
-{\grad\cdot \Fvec} + q = \rho C_p \frac{\di T}{\di t},
\end{equation}
where $q$ is the rate of energy release by an internal heat source within the ``lumped system'', $\rho$ is the bulk density 
and $C_p$ is the heat capacity. 

\subsubsection{Amorphous to crystalline ice transition}\label{SSS:amcr}
The crystallization process of amorphous ice is spontaneous but greatly affected by temperature.
Its rate is given by \citep{Schmitt1989}, based on laboratory experiments
\begin{equation}\label{eq:crstlrate}
\lambda=1.05\times 10^{13}\,\text{e}^{-5370/T}\quad \left[s^{-1}\right]
\end{equation}
The amorphous ice mass fraction $X_{A}$ changes with time at a rate
\begin{equation}
\dot{X}_{A}=-\lambda \left( T \right) X_{A}
\end{equation}
This process, which changes the ice structure, also changes its physical properties, such as density and thermal 
conductivity. In addition, crystallization is an exothermic process that can cause chain-reaction, that is, propagate
by feeding on its own energy.
The rate of heat generated by crystallization is given by
\begin{equation}
q=\lambda \left( T \right) \rho X_{A} \mathcal{H}_{a}
\end{equation}
where $\mathcal{H}_{a}$ is the energy released per unit mass \citep{Ghormley1968}.

\subsection{Boundary conditions}\label{SS:boundary}
Boundary conditions at the surface consist of balance of heat fluxes into the nucleus from external sources and out of it into space.
Three such sources are considered:
\begin{enumerate}
\item Solar radiation flux, given by 
\begin{equation}
(1-\mathcal{A})\frac{ L_{\odot}}{4\pi d_{H}^{2}}\cos{\xi},
\end{equation}
where $\mathcal{A}$ is the surface albedo, $L_{\odot}$ is the solar luminosity, $d_{H}$ is the heliocentric 
distance and $\xi$ is the local solar zenith angle. Here
$\cos{\xi}$ is calculated for surface coordinates $\theta$ and $\phi$ according to
\begin{equation}
\cos{\xi}=\sin{\theta_s}\sin{\theta}\cos{\left(\phi_s+\phi\right)}+\cos{\theta_s}\cos{\theta},
\end{equation}
where $\theta_s$ and $\phi_s$ are the sub-solar coordinates.
\item Thermal radiation loss, given by
\begin{equation}
\epsilon \sigma T^{4},
\end{equation}
where $\epsilon$ is the emissivity and $\sigma$ is the Stefan-Boltzmann constant.
\item Latent heat absorbed per unit time by sublimation of volatiles, given by 
\begin{equation}\label{eq:sublimation}
\mathcal{P}_v\left(T\right)\sqrt{\frac{m}{2\pi kT}}H,
\end{equation}
where  $m$ is the molecular weight, $H$ is the latent heat of sublimation and $\mathcal{P}_v$ is the vapor pressure, given by the Clausius-Clapeyron equation
\begin{equation}
\mathcal{P}_v=a\text{e}^{-b/T}
\end{equation}
\end{enumerate}

\subsection{Numerical scheme}\label{SS:numeric}
The heat equation (\ref{eq:heat}) is a non-linear second-order partial differential equation, which has no analytical solution and has
to be solved numerically.
The differential equation is transformed into a set of difference equations, and applied to a finite grid, where
the infinitesimals $\delta R, \delta\Theta, \delta\Phi$ and $\delta t$ become the grid steps 
$\Delta R, \Delta\Theta, \Delta\Phi$ and $\Delta t$, respectively.

\begin{figure}
\centering
\includegraphics[scale=0.55]{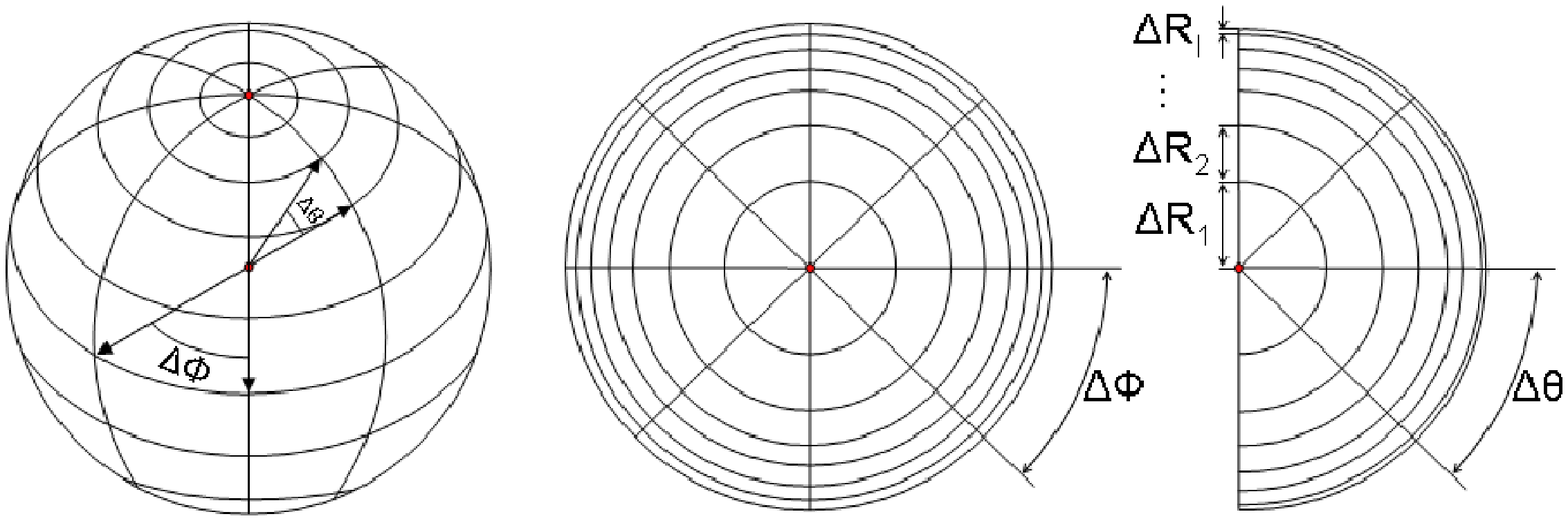}
\caption{Logical representation of the numerical grid, conforming to spherical coordinates convention. 
The $\Theta$ coordinate is divided into '$J$' equal intervals ($\Delta\Theta$), the $\Phi$ coordinate is divided 
into '$K$' equal intervals ($\Delta\Phi$) and the radial coordinate is divided with geometric progression into '$I$' intervals 
($\Delta R_i$). The equations of the numerical scheme are based on this representation.}
\label{fig:logicalgrid}
\end{figure}

The grid has two representations, the logical (geometric) one and the computerized (algebraic) one.
The logical grid, shown in Fig.~\ref{fig:logicalgrid}, is the spatial representation of the 
difference equations, and is a 3-D grid in spherical coordinates.
The grid itself is a 3-D mesh with $I$,$J$ and $K$ divisions for $R$ (radial distance), 
$\Theta$ (co-latitude) and $\Phi$ (azimuth).
In order to better study surface phenomena, $\Delta R$ is taken to be decremental geometrically, and is uniquely
determined by the radius of the nucleus, the number of divisions (`$I$') and the surface layer's thickness. 
Dimensions $\Theta$ and $\Phi$ are divided into equal intervals.

\begin{figure}
\centering
\includegraphics[scale=0.55]{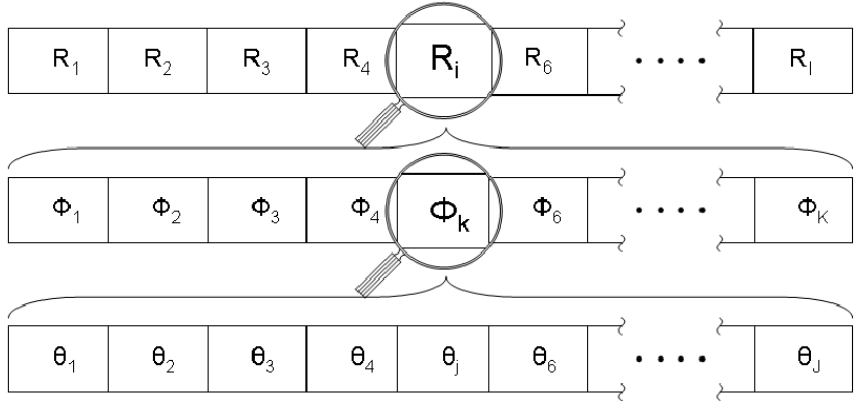}
\caption{Computerized (vector) representation of the numerical grid. This representation is the order of the grid's elements in the solution vector of size $n\equiv I\!\cdot\!J\!\cdot\!K$. The relation between the logical and vector grids is given by equation (\ref{EQ:grids}). In this representation $R$ varies the slowest through the vector and $\Theta$ varies the fastest. The equations matrix, according to this representation, will be diagonal (albeit not tight) and easier so solve.}
\label{fig:compgrid}
\end{figure}

The computerized grid is a vector with $n\equiv I\!\cdot\!J\!\cdot\!K$ elements. The elements are 
organized along the vector, as shown in Fig.~\ref{fig:compgrid}, so that radius values will vary the slowest, 
and the co-latitude will vary the fastest. 
The relation between the logical and the computerized grid is expressed by:
\begin{equation}\label{EQ:grids}
v=j+\left(k-1\right)J+\left(i-1\right)JK
\end{equation}
$v$ is the position in the computer grid vector, while $i$,$j$ and $k$ are the indexes for $R$,$\Theta$ and $\Phi$ and
$J$ and $K$ are number of divisions for $\Theta$ and $\Phi$.
This configuration forms a diagonal equation matrix, which is easier to solve.

\subsection{Difference equations}\label{SS:differences}
For greater stability, and in order to remove time step constraints, a fully implicit iterative scheme is chosen,
although the formulation is more complicated and has a higher computational load.
The difference equation for each volume element has the form
\begin{equation}\label{EQ:heat}
\sum{F_{x}^{(n)}\Delta S_x}+q^{(n)}\Delta V=\frac{U^{(n)}-U^{(0)}}{\Delta t} \Delta V
\end{equation}
where $x$ is the direction ($i^+$,$i^-$,$j^+$,$j^-$,$k^+$ and $k^-$), $n$ is the iteration number, $F_x$ is the heat flux
vector component in the $x$ direction,
 $\Delta S_x$ is the element's $x$-side area, $q^n$ is produced heat inside the element, $\Delta V$ is the element's volume,
 $\Delta t$ is the time step, and $U$ is the thermal energy.
We note that the difference scheme is conservative, that is, upon integration over the entire volume, all fluxes cancel out
except the flux crossing the nucleus surface. In other words, Gauss's theorem is satisfied by the difference
equations on the discrete grid. The sum on the LHS of equation (\ref{EQ:heat}) runs over all sides of a volume element: 6 for most elements, 5 for those ending on the $\Theta=0$ or $\pi$ axis, and 4, for those ending at the center $R=0$.

The equations are linearized and 
solved for $\Delta T$ rather than for $T$ itself; as $\Delta T$ reduces over iterations, it serves as convergence 
criterion.

\section{Tests of the 3-D model}\label{S:tests}
\subsection{Analytical and numerical tests}\label{SS:analyt}
Several comparative tests were carried out to ensure that heat flow is correctly solved.
First, two analytical solutions for the heat transport equation on a sphere were chosen. 
Since all terms in our surface boundary condition (solar radiation, sublimation and thermal emission) are non-linear, 
and hence do not allow analytical solutions to the heat equation, we chose a linear convective boundary condition. 
The numerical model was modified to the same boundary condition, and kept spherically symmetric in order to become
comparable to the analytical solution.
The first analytical model (hereafter, C\&J) was taken from the literature \citep{CNJ}:

\begin{equation}
T\left(r,t\right)=\frac{2hT_0}{r}\sum_n \frac{\sin\left(\lambda_na\right)\left[\left(\lambda_na\right)^2+\left(1-ah\right)^2\right]}
{ \lambda_n^2\left[\lambda_n^2-ah\left(1-ah\right)\right] }\sin\left(\lambda_nr\right)e^{-\lambda_n^2Dt}
\end{equation}
$\lambda_n$ defined as:
\begin{equation}\label{EQ:lambda}
\tan\left(\lambda_na\right)=\frac{\lambda_na}{1-ah}
\end{equation}
where $T_0$ is the initial temperature, $a$ is the sphere's radius, $h$ is the convection constant, $D$ is 
the conduction constant and $t$ is the time step; $\lambda_n$ is the series solution of Eq.(\ref{EQ:lambda}).
This solution is based on simplifying assumptions.

A second, accurate analytical solution was derived:
\begin{equation}
T\left(r,t\right)=\frac{2T_0}{r}\sum_n \frac{\sin\left(\lambda_na\right)-\left(\lambda_na\right)\cos\left(\lambda_na\right)}
{ \lambda_n^2a-\lambda_n\sin\left(\lambda_na\right)\cos\left(\lambda_na\right) }\sin\left(\lambda_nr\right)e^{-\lambda_n^2Dt}
\end{equation}
which is independent of $h$, and therefore not limited by boundary condition values.

A different test model was obtained by solving the heat equation numerically by a 1-D spherical explicit scheme:
\begin{equation}
T_i^{(n+1)}=T_i^{(n)}+2D\frac{\Delta t}{\Delta V_i}\left[\frac{T_{i-1}^{(n)}-T_i^{(n)}}{\Delta r_i+\Delta r_{i-1}}S_{i}+\frac{T_{i+1}^{(n)}-T_i^{(n)}}{\Delta r_i+\Delta r_{i+1}}S_{i+1}\right]
\end{equation}
with boundary condition ($i=I$):
\begin{equation}
T_I^{(n+1)}=T_I^{(n)}+2D\frac{\Delta t}{\Delta V_I}\left[\frac{T_{I-1}^{(n)}-T_I^{(n)}}{\Delta r_I+\Delta r_{I-1}}S_{I}-\frac{1}{2}hT_s^{(n)}S_{I+1}\right]
\end{equation}
where
\begin{equation}
\Delta V_i=\frac{4}{3}\pi \left[\left(R_i+dR_i\right)^3-R_i^3\right]
\end{equation}
and
\begin{equation}
S_i=4\pi R_i^2
\end{equation}

\begin{figure}
\centering
\includegraphics[scale=0.5]{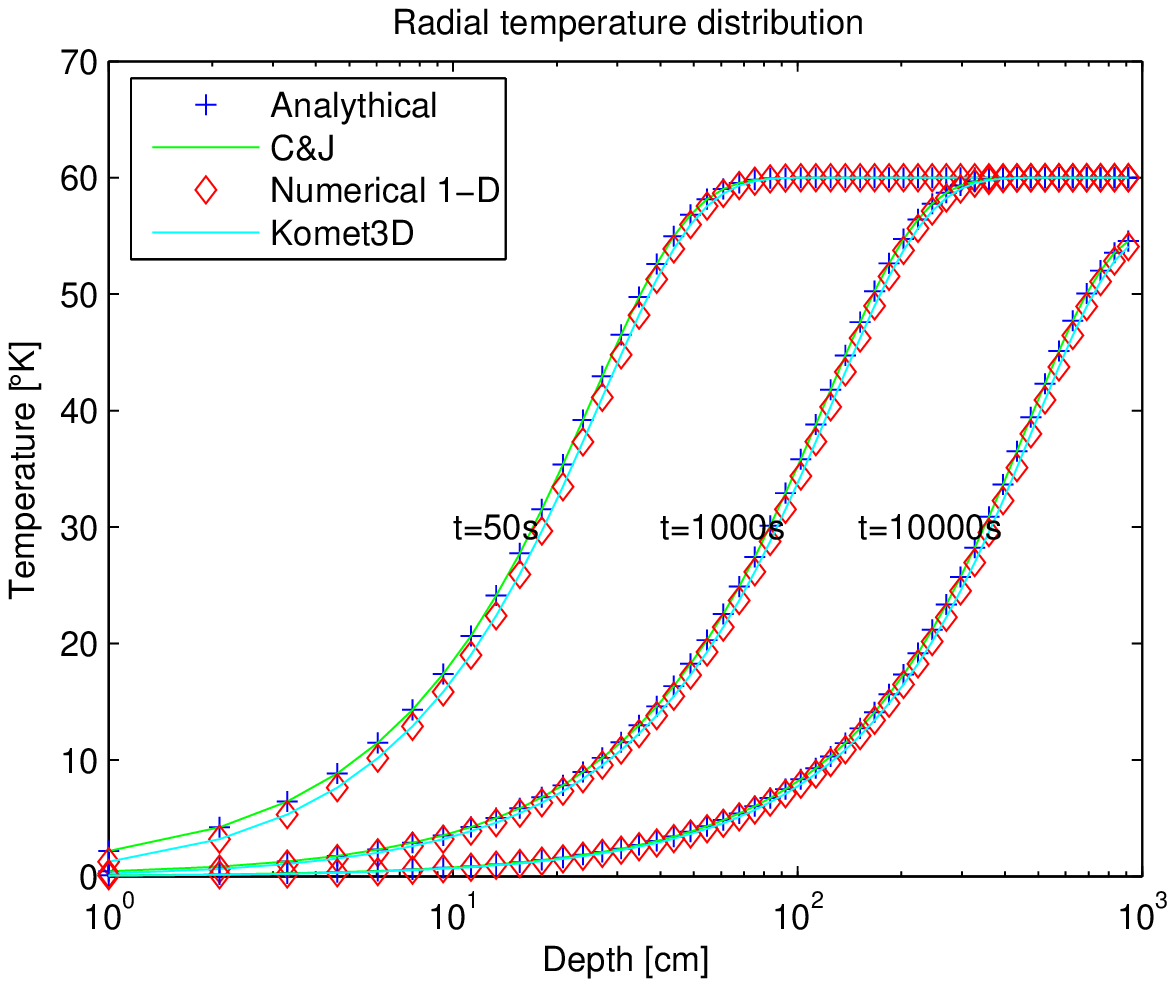}
\includegraphics[scale=0.5]{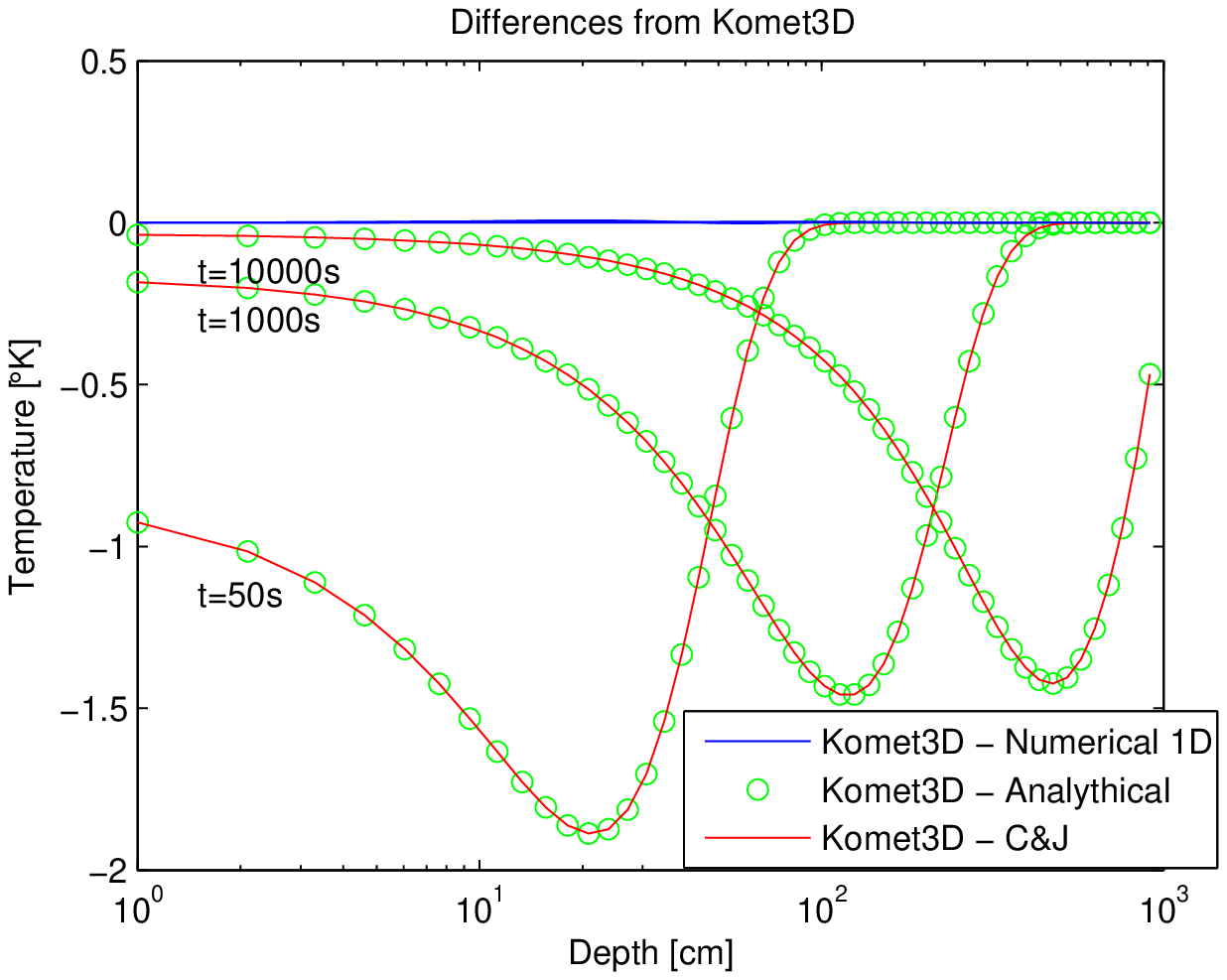}
\caption{Different solutions for a cooling sphere with uniform initial conditions, 
and linear convection boundary condition, at 50, 1000 \& 10000 seconds. All solutions: Carslaw \& Jaegar, analytical development, 
numerical 1-D and the 3-D (Komet3D) models converge to the same result ({\it left}). The absolute difference between all models 
to the 3-D model ({\it right}) reveals perfect agreement between the numerical models, and up to 5\% error for the analytical solutions. }
\label{fig:models}
\end{figure}

The tests were performed on a sphere with $R=1000$~cm. All other parameters were chosen artificially 
(to allow the C\&J model
to converge) and do not represent a comet nucleus. The sphere was initially set at a uniform temperature of $60^{\circ}K$
and was then allowed to cool. 
Temperature profiles as function of depth are shown in Fig.~\ref{fig:models} at three different times for all 4 solutions.
Clearly, the 3-D model agrees very well with both analytical models, as well as with the 1-D numerical model at all
times. A more detailed comparison of the model with the other models is obtained by plotting the differences between them,
as shown in the right panel of Fig.~\ref{fig:models}.
The maximal difference between the 3-D model and the analytical models is less than 5\%, which is an excellent agreement, keeping in mind that the lowest radial grid resolution is 9\%; the difference between the two numerical models is 
negligibly small.

\subsection{Comparison with other models for comet 46P/Wirtanen}\label{SS:wirtanen}

\begin{figure}
\centering
\includegraphics[scale=0.38]{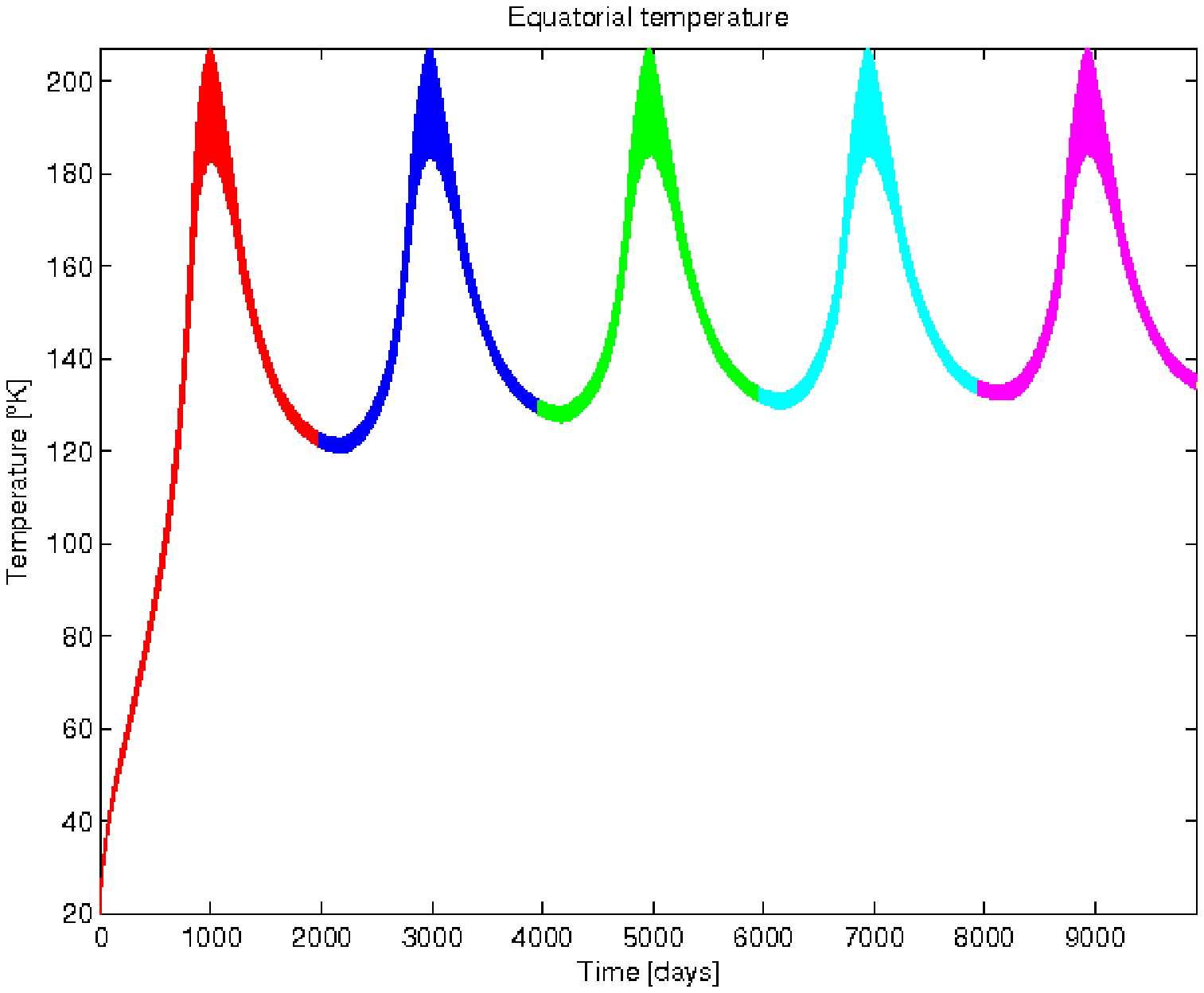}
\includegraphics[scale=0.38]{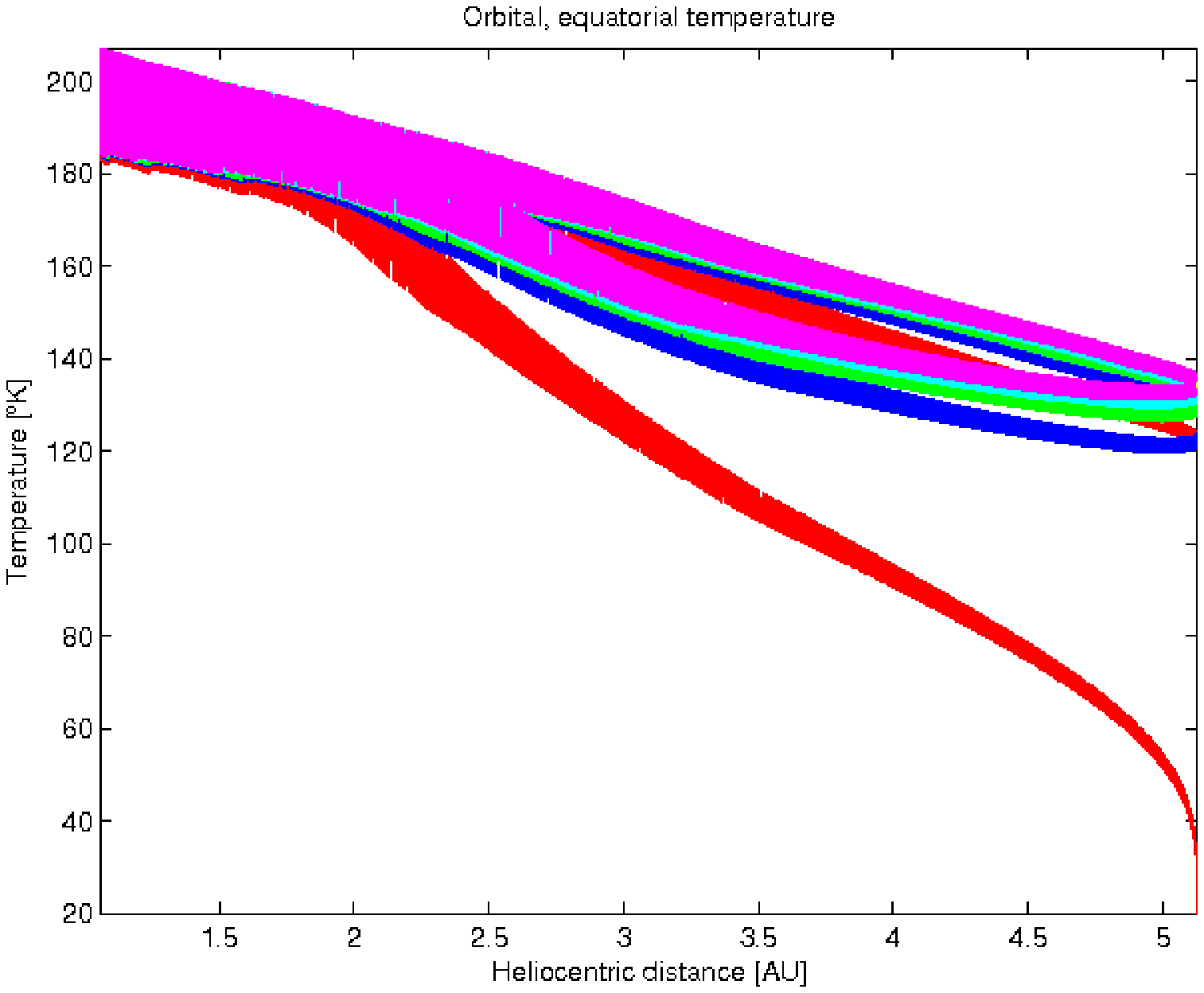}
\caption{Orbital temperature variation of an equatorial model element with comet Wirtanen's orbital parameters, 
for 5 revolutions around the sun. Colors distinguish between revolutions. Line thickness indicates temperature differences 
between day and night sides.}
\label{fig:K3D-W-temp}
\end{figure}

Comet 46P/Wirtanen was intensively studied while it was the target of space mission {\it Rosetta}. In particular,
several different and independent numerical codes were applied to this comet and the results were carefully
analyzed and compared (Huebner et al. 1999). These models, allowing for differences among them, can be considered
a reliable test for the present model. We therefore used our code adopting comet 46P/Wirtanen's parameters
(a=3.093738~AU, e=0.6578222, R=600m, and a spin period of 24~hr, as used in other model calculations), and
calculated the thermal evolution for 5 revolution around the sun, assuming the spin axis to be perpendicular to the
orbital plane (zero tilt angle).  The nucleus composition consisted of 50\% 
crystalline H$_2$O ice and 50\% dust, by mass. No Hertz factor was used ($H_z=1$).
In Fig.~\ref{fig:K3D-W-temp}, the temperature of a point on the equator is shown versus time and heliocentric distance. 
The line thickness represents daily temperature variations. As expected, the perihelion (daytime) temperature 
is the same for all orbits, since heat exchange with the interior is negligible in this case, and the surface
temperature is controlled by sublimation. 
The aphelion temperature increases slightly with each period; again, this is to be expected, as the nucleus 
surface layer accumulates heat and needs several revolutions to stabilize. The results are in very good agreement
with Huebner et al. (loc. cit.), for a model with similar parameters. We also compared our results with the quasi-3-D
model of comet 46P/Wirtanen computed by \citep{Cohen2003} and found very good matching of the temperature map.

\begin{figure}
\centering
\includegraphics[scale=0.52]{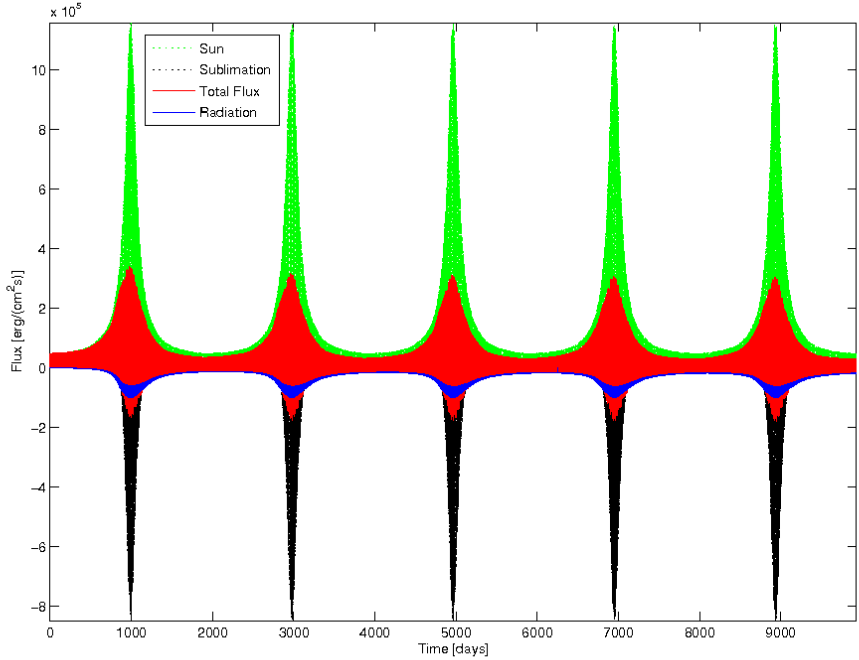}
\caption{Surface flux balance for equatorial element, with comet Wirtanen's orbital parameters 
(same as in Fig.~\ref{fig:K3D-W-temp}). Positive values are inward fluxes, and negative are outwards fluxes 
(relative to the nucleus center). The red line represents total flux.}
\label{fig:K3D-W-flux}
\end{figure}

Fig.~\ref{fig:K3D-W-flux} illustrates boundary fluxes for that same point on the equator. 
Positive fluxes are directed inward, and negative 
fluxes outward. At perihelion, the solar flux is mostly absorbed by sublimation, while at aphelion the various terms are comparable. The residual flux, which represent heat conducted into or out of the nucleus, shows that on the
day side, there is a net inward flux, and on the night side, a net outward flux.

\section{Applications of the 3-D code}\label{S:Results}

\subsection{Models of Comet 67P/ Churyumov-Gerasimenko}\label{SS:CG}

\begin{figure}
\centering
\includegraphics[scale=0.455]{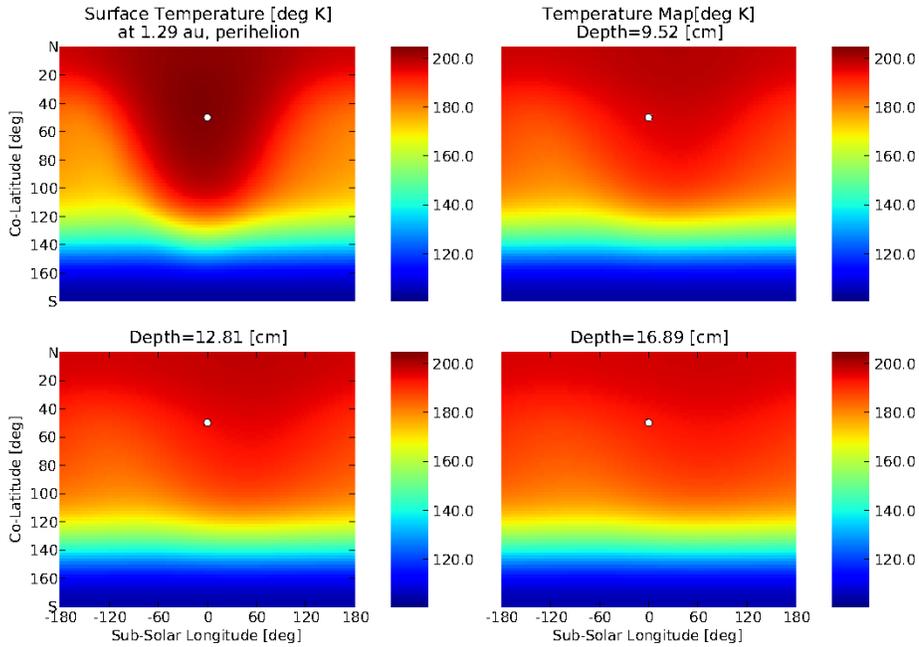}
\caption{Shell temperatures for different depths for comet 67P/C-G 
with 40$^\circ$ tilt angle. The theoretical daily skin-depth for 50\% crystalline water and 50\% dust 
with Hertz factor of $H_z=0.1$, was calculated to be 11~cm. The maps indicates that the azimuthal 
temperature homogeneity increases with depth, and that deeper than 16~cm, hourly temperatures are almost indistinguishable.}
\label{fig:skindepthdaily}
\end{figure}

\begin{figure}
\centering
\includegraphics[scale=0.455]{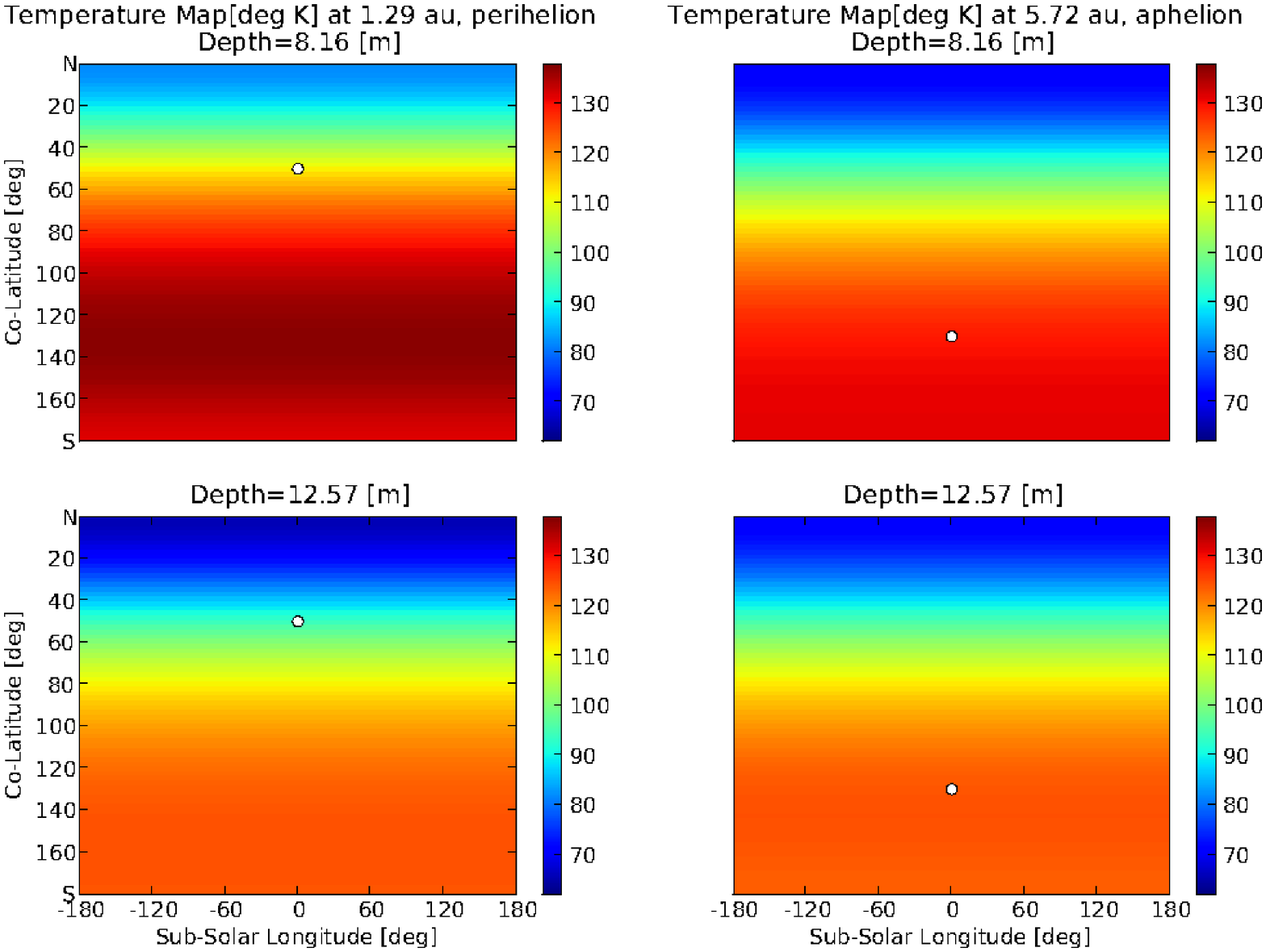}
\caption{Orbital skin-depth for comet 67P/C-G 
with 40$^\circ$ tilt angle. For reference, the analytical value for 50\% crystalline water and 50\% dust 
with $H_z=0.1$ is 8~m. Top maps show a spherical shell, 8~m deep, at perihelion 
({\it left}) and at aphelion ({\it right}). Differences between them indicate that this depth is within the orbital 
skin-depth. The bottom maps show temperatures of a deeper shell (12.5~m), where differences are minute, indicating 
that the shell is deeper than the orbital skin-depth.}
\label{fig:skindepthannual}
\end{figure}

The actual target of the {\it Rosetta} mission is now comet 67P/ Churyumov-Gerasimenko (hereafter, 67P/C-G). We have used
our code to model the evolution of a spherical nucleus that has the characteristics of this comet,  a=3.5029497~AU, e=0.6319359, R=1980m, and a spin period of 12.6~hr \citep{Lamy2004}, in particular 
a spin axis tilt of 40$^\circ$-45$^\circ$ \citep{Chesley2004}.
We considered a nucleus containing 50\% water crystalline ice and 50\% dust by mass, and  calculated the 
thermal evolution for 5 orbital revolutions. Figures~\ref{fig:skindepthdaily} and \ref{fig:skindepthannual} illustrate 
the daily and orbital skin depths, by comparing several temperature maps for different depths and orbital positions. 
Analytical diurnal and orbital skin-depths were calculated to be 11~cm and 8~m respectively, in good agreement with the 
numerical results.

\begin{figure}
\centering
\includegraphics[scale=0.45]{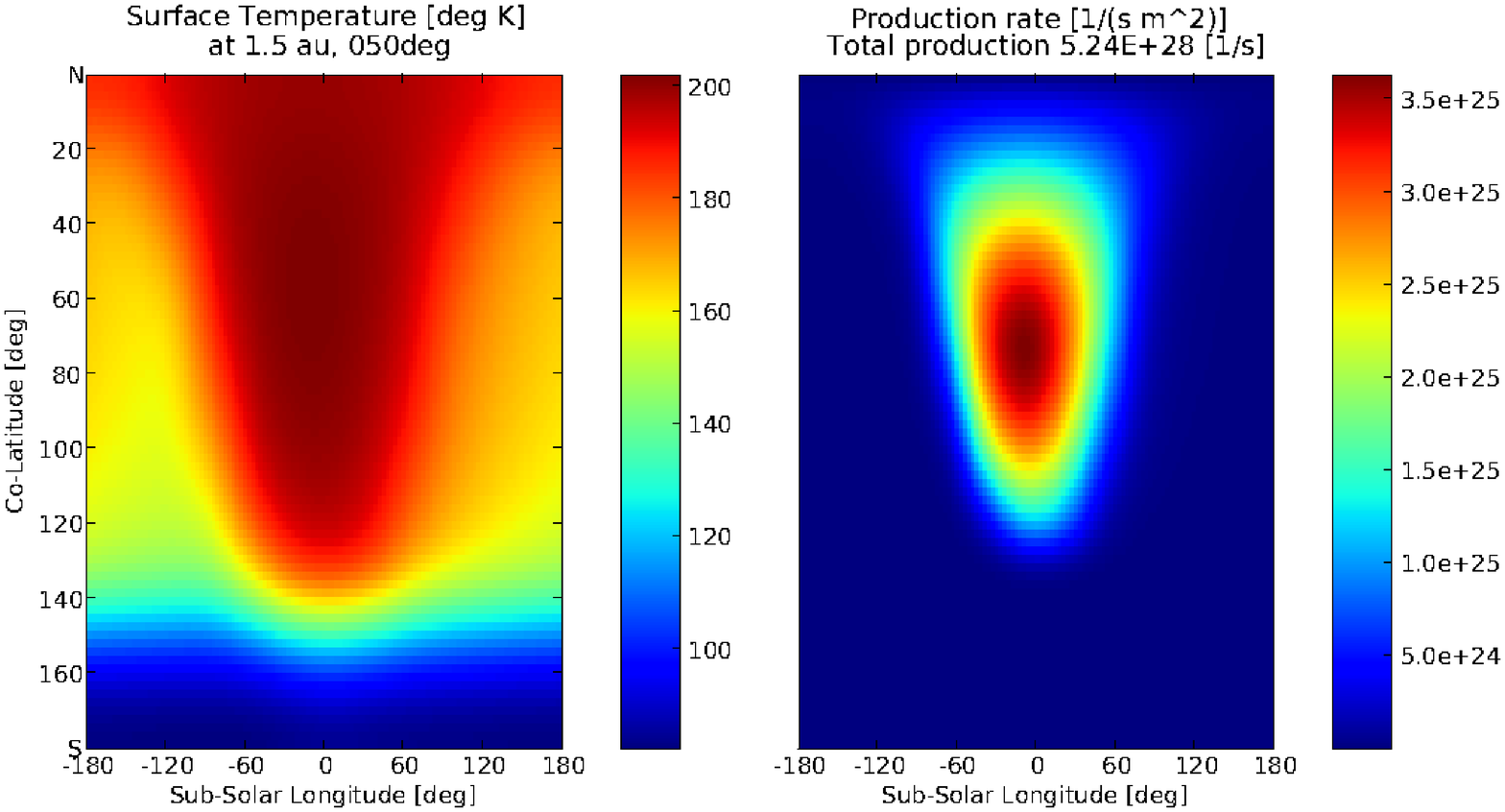}
\includegraphics[scale=0.45]{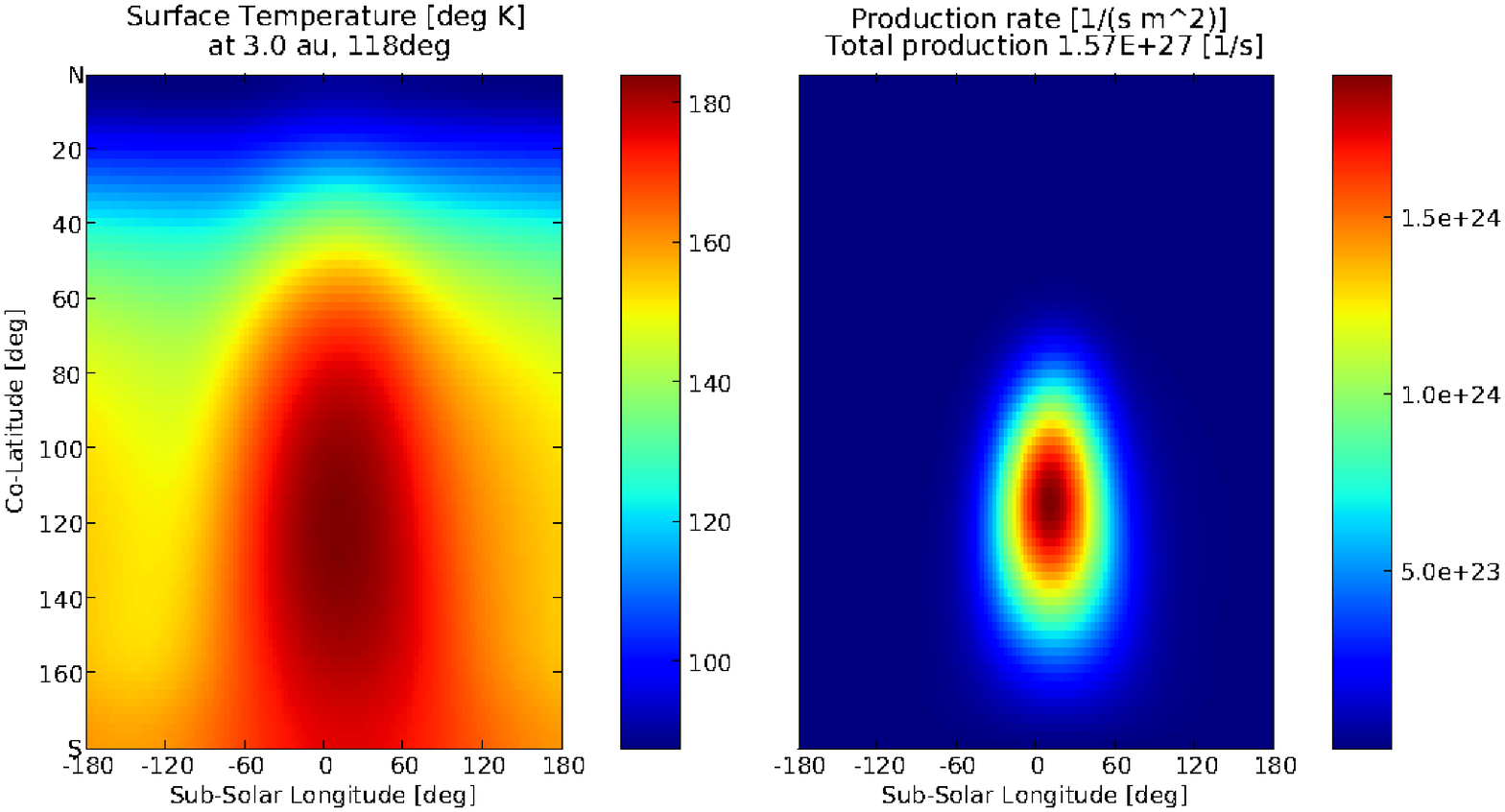}
\caption{Surface temperature and water production rate maps for comet 67P/C-G 
with 45$^\circ$ tilt angle, where crystallization is taken into account. Top maps show temperature ({\it left}) and 
local flux (m$^{-2}$~s$^{-1}$) of water molecules ({\it right}) at 1.5~AU post-perihelion, and bottom maps, at 3~AU post-perihelion.}
\label{fig:production}
\end{figure}

Another model was calculated assuming an initial composition of 50\% amorphous ice and 50\% dust by mass and
allowing for crystallization of the amorphous ice.
The water production rate over the entire nucleus surface is shown in Fig.~\ref{fig:production} together with the
respective temperature map at two heliocentric distances post-perihelion. The strong temperature dependence of the
sublimation rate is illustrated by the high concentration of the active spot. We note that the peak is slightly
shifted beyond the subsolar point (toward afternoon).
The maximal integrated total production rate obtained (near perihelion), $8.3\!\cdot\!10^{28}$~s$^{-1}]$, is 
higher than the observed value of $\sim1\!\cdot\!10^{28}$~s$^{-1}$
\citep{Schleicher2003}; this is not surprising
since the model assumes water sublimation from the entire surface, while active areas are most probably
confined to a fraction of the
nucleus surface, as indicated by close-up observations of comet nucleus surfaces \citep{Keller1986,Sunshine2006}.

\subsection{The effect of the spin-axis tilt}\label{SS:AxisTilt}

Variations of an input parameter may lead to different behavior patterns for models that otherwise have the 
same properties. 
Here we show, for the first time, the effect of the spin-axis tilt on the temperature distribution 
over the surface of a short-period comet. For illustration, we have chosen the orbital parameters of comet 
67P/ C-G. Our aim is to determine the minimal and maximal temperatures that can be obtained locally, 
under all possible inclinations. This will shed light on the viability of very volatile ices on the surface of comet nuclei.

Assuming a spherical comet nucleus with the orbital parameters of comet 67P/C-G, we carried out evolutionary calculations
over the entire range of tilt angles, at intervals of 10$^\circ$. For all models, the summer solstice point was taken to 
be at perihelion, (i.e. the projection of the spin vector on the ecliptic plane, points to the sun at perihelion). 
Thermal evolution was calculated over a period of 5 orbital revolution. The nucleus composition was taken to be 
50\% dust and 50\% crystalline H$_2$O ice by mass. Since we were interested in 
the temperature distribution at the surface, and since for a short-period comet, amorphous ice is present only
at relatively large depths, we have assumed crystalline ice throughout. We adopted a 0.1 Hertz factor.

\begin{figure}
\centering
\includegraphics[scale=0.21]{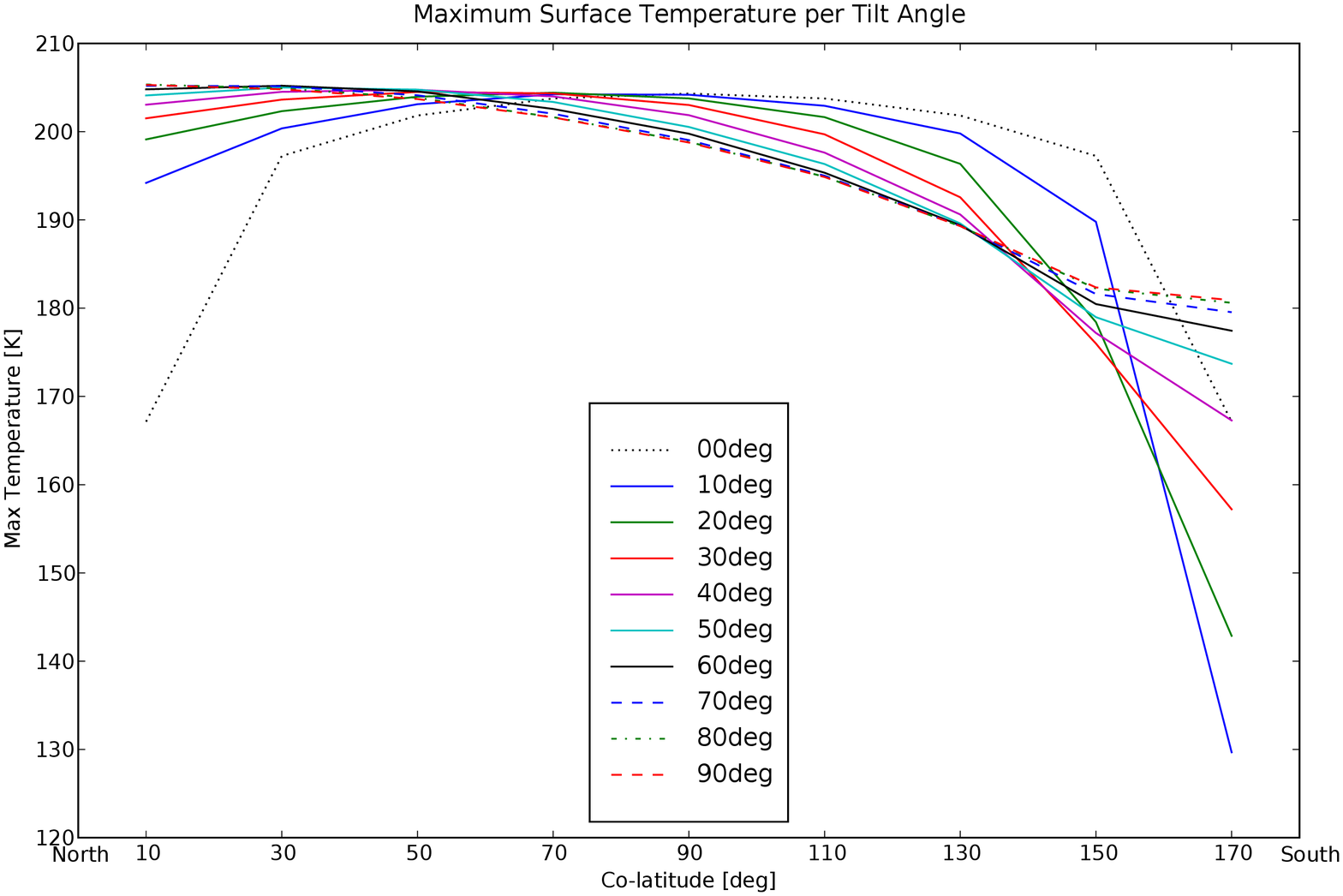}
\includegraphics[scale=0.21]{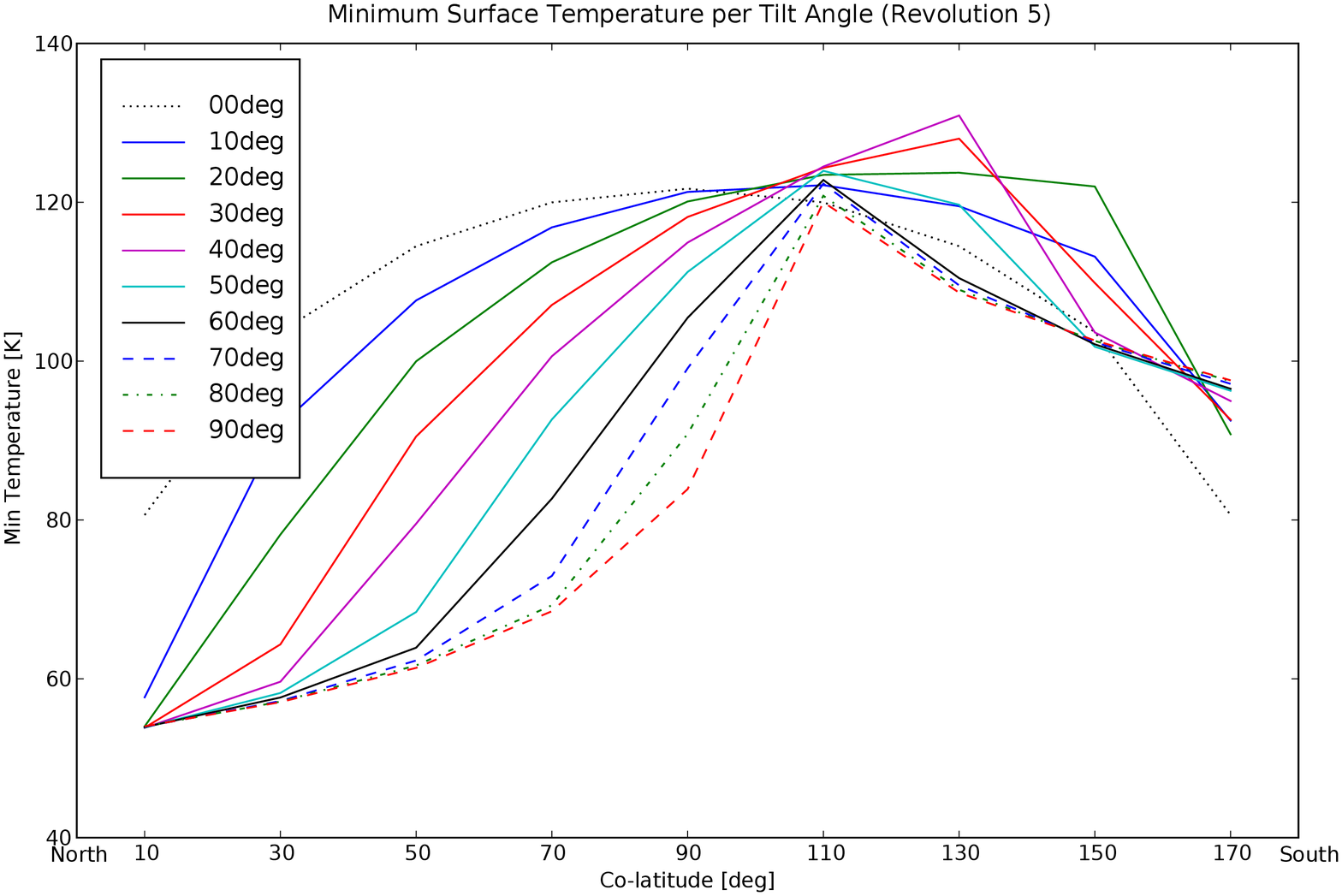}
\caption{Maximum and minimum local surface temperatures for 10 different spin-axis tilts. Extremum temperatures vary 
only with co-latitude, and are greatly affected by tilt angle. The lowest value of the maximum temperature is obtained
at the southern pole for small tilt angle. The lowest temperatures (over the $5^{th}$ revolution) are found at the north pole, 
where the highest maximum temperature occurs.}
\label{fig:tilt}
\end{figure}

The results are summarized in Fig.~\ref{fig:tilt}. For maximum temperatures (left), south-pole temperatures 
were always lower than the north-pole ones; however, for small tilt angles (below 40$^\circ$) the difference 
was found to be more significant. When the tilted nucleus is at perihelion, the south-pole is directed away 
from the sun, not receiving any solar flux. The only heat flowing into the south pole is from non-lateral 
conduction, which is negligible relative to the solar flux. The south-pole remains hidden from solar flux 
until the nucleus nears the equinox, where the sun is directly above the nucleus' equator, and the south 
pole is exposed to the sun at a very low angle. Passing the equinox point, the south pole is exposed to direct flux. 
As the comet retreats from equinox, the flux received by the ever exposed south-pole depends on the tilt angle, 
making the small tilt angle pole receive only a fraction of the flux received by a high tilt pole.
The lowest value of the maximal surface temperature is less than 130$^\circ$K. At such temperatures crystallization of amorphous 
water ice proceeds very slowly: the characteristic timescale is of the order of days, competing with the
dynamical (orbital) timescale. It is thus possible that amorphous ice be retained at (or very nearly below) the surface
of a very confined area, provided that the tilt angle remains unchanged. On the other hand, this temperature is
far too high for ices of volatile substances (certainly CO, CH$_4$, N$_2$, but also CO$_2$, NH$_3$, HCN) to survive 
insolation at the surface. These molecules --- observed to be ejected by comet nuclei --- must, 
therefore originate in deeper layers or 
be released by crystallization of gas-laden amorphous water ice. 
It is, therefore, inconceivable that pristine material be found on the surface of comets of the Jupiter family type.

Minimum temperatures (right panel of Fig.~\ref{fig:tilt}) are lowest for the north pole, which is allowed to 
cool for most of the comet's orbit. 
Temperatures drop below $60^\circ$K, which may render the surface and the outer layers almost completely inert.

\begin{figure}
\centering
\includegraphics[scale=0.54]{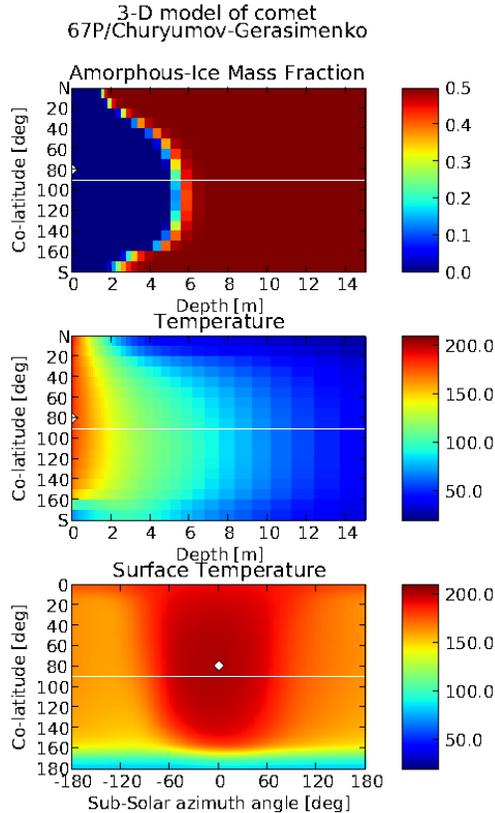}
\caption{Surface temperature and advance of the crystallization front 
for comet 67P/C-G with 15$^\circ$ tilt angle at 1.5~AU post-perihelion. 
The model was started with 50\% dust and 50\% amorphous ice with $H_z=0.01$ and the amorphous ice was 
allowed to crystallize. Top panels show the amorphous ice mass fraction 
and temperature as function of co-latitude angle and depth, with arbitrary azimuth. The crossing white line  
represents the equatorial plane and the white dot at the ordinate represents the sub-solar co-latitude angle. 
Surface temperature ({\it bottom}) is shown with same equatorial and subsolar markings.}
\label{fig:crystallization}
\end{figure}

To further study the influence of small tilt angles on sub-surface crystallization, a high-resolution model of
comet 67P/C-G was computed, adopting a 15$^\circ$ tilt angle. This high-resolution run of the model took 3 months of calculations on a 3.0Ghz 64bit Pentium-4 based PC. Therefore, it was not conducted exhaustively for each tilt angle.
The model started with 50\% amorphous ice that 
was allowed to crystallize. Fig.~\ref{fig:crystallization} shows the crystallization cross-section pattern (top) 
and the temperatures for the same area (middle) at 1.5~AU post-perihelion. It is shown that the south-pole 
crystallization front did not exceed depth of 3m (during the entire run of the model the front did not advance 
beyond 4m), in contrast to a model with 40$^\circ$ tilt angle (not shown here) that reached a depth of 10~m 
within the same period. The north-pole crystallization front did not exceed the depth of 2~m in both models, 
although maximum temperatures were the highest. This happened due to the rapid exposure of the pole to the solar flux. 
When the pole is hidden from the sun (most of the time), the cooling boundary conditions with only small amount 
of heat that has been absorbed during solar exposure, prevent the deeper temperature from increasing, and thus 
crystallization propagates only to a shallow depth. This may indicate that pristine material may be found at 
the north-pole at small depths but not at the surface.

We note that our present model does not account for 
possible recession of the surface due to sublimation of the ice. This effect would distort the sphericity of the
model (numerical grid). On the other hand, as stated above, most of the surface of comet nuclei is covered by
a dust mantle, presumably of high porosity, and water vapor is produced and ejected from an ice-rich layer lying beneath
the dust mantle. Therefore, the process of erosion may be more complicated than just simple ablation and requires
separate investigation. The point we wish to stress in the present calculation is that amorphous ice may lie very
close to the nucleus surface under appropriate conditions, although it would not be detectable on the surface itself.

\section{Summary and Conclusions}\label{S:Summary}

We have developed a fully 3-D thermal evolution code for comet nuclei that may, in fact, be applied to any small body
of the solar system, small enough for self-gravity to be negligible. In the present paper, which is the first of a series,
we have tested the code with respect to analytical solutions, as well as 1-D and quasi-3-D comet nucleus models, and found
excellent agreement.

We computed models of comet 67P/C-G, adopting a 1:1 mass ratio of ice to dust. In this case, surface temperatures
are mainly determined by the ice, through the strongly temperature-dependent sublimation term of the boundary condition.
Assuming a spin axis tilt of 40$^\circ$, we obtained a maximum surface temperature of 205$^\circ$K and a minimum temperature of 54$^\circ$K.

We have investigated the effect of the spin axis tilt on the surface temperature distribution and found that 
conditions for preservation of pristine amorphous ice and moderately volatile species at shallow depths is possible 
for low tilt angles, when some surface areas are shielded from insolation. This is despite lateral heat 
conduction driven by the strong temperature variations at the nucleus surface. Although not shown here, more distant 
comets with low tilt angles could retain unprocessed surface amorphous ice. 

A fully 3-D comet nucleus model opens a wide range of possible subjects of study, involving inhomogeneities, 
both innate and evolutionary, which appear to be so common to comets. We shall address some of them in upcoming papers. 

\textbf{Acknowledgments}
Support for this work was provided by the Israel Science Foundation grant No. 942/04.

\end{document}